\documentclass[prb,amsmath,amssymb,floatfix, 12pt]{revtex4}
\usepackage{graphicx}
\usepackage{bm}
\begin{document}
\def \tr{{\mbox{tr~}}}
\def \ra{{\rightarrow}}
\def \ua{{\uparrow}}
\def \da{{\downarrow}}
\def \be{\begin{equation}}
\def \ee{\end{equation}}
\def \ba{\begin{array}}
\def \ea{\end{array}}
\def \bea{\begin{eqnarray}}
\def \eea{\end{eqnarray}}
\def \nn{\nonumber}
\def \l{\left}
\def \r{\right}
\def \half{{1\over 2}}
\def \etal{{\it {et al}}}
\def \cH{{\cal{H}}}
\def \cM{{\cal{M}}}
\def \cN{{\cal{N}}}
\def \cQ{{\cal Q}}
\def \cI{{\cal I}}
\def \cV{{\cal V}}
\def \cG{{\cal G}}
\def \bS{{\bf S}}
\def \bI{{\bf I}}
\def \bL{{\bf L}}
\def \bG{{\bf G}}
\def \bQ{{\bf Q}}
\def \bR{{\bf R}}
\def \br{{\bf r}}
\def \bu{{\bf u}}
\def \bq{{\bf q}}
\def \bk{{\bf k}}
\def \bz{{\bf z}}
\def \bx{{\bf x}}
\def \tJ{{\tilde{J}}}
\def \W{{\Omega}}
\def \e{{\epsilon}}
\def \lam{{\lambda}}
\def \L{{\Lambda}}
\def \a{{\alpha}}
\def \t{{\theta}}
\def \b{{\beta}}
\def \g{{\gamma}}
\def \D{{\Delta}}
\def \d{{\delta}}
\def \w{{\omega}}
\def \s{{\sigma}}
\def \f{{\varphi}}
\def \x{{\chi}}
\def \e{{\epsilon}}
\def \h{{\eta}}
\def \G{{\Gamma}}
\def \z{{\zeta}}
\def \hatt{{\hat{\t}}}
\def \hn{{\bar{n}}}
\def \vk{{\bf{k}}}
\def \vq{{\bf{q}}}
\def \gk{{\g_{\vk}}}
\def \nd{{^{\vphantom{\dagger}}}}
\def \yd{^\dagger}
\def \av#1{{\langle#1\rangle}}
\def \ket#1{{\,|\,#1\,\rangle\,}}
\def \bra#1{{\,\langle\,#1\,|\,}}
\def \braket#1#2{{\,\langle\,#1\,|\,#2\,\rangle\,}}

\title{Decay of super-currents in condensates in optical lattices}
\author {Anatoli~Polkovnikov, Ehud~Altman, Eugene~Demler, Bertrand~I.~Halperin, and Mikhail~D.~Lukin}
\address{Physics Department, Harvard University, Cambridge, MA 02138}
\date{\today}
\begin{abstract}
In this paper we discuss decay of superfluid currents in boson
lattice systems due to quantum tunneling and thermal activation
mechanisms. We derive asymptotic expressions for the decay rate
near the critical current in two regimes, deep in the superfluid
phase and close to the superfluid-Mott insulator transition. The
broadening of the transition at the critical current due to these
decay mechanisms is more pronounced at lower dimensions. We also
find that the crossover temperature below which quantum decay
dominates is experimentally accessible in most cases. Finally, we
discuss the dynamics of the current decay and point out the
difference between low and high currents.

\end{abstract}

 \maketitle

\section{introduction}

Some of the most intriguing questions in low temperature physics concern
the ways in which superconductors lose their superconducting
properties, because of thermal or quantum fluctuations.  Mike Tinkham has long
been fascinated with these issues, and has done much to advance our
understanding of the subject.

An early contribution in this area was the work of Newbower,
Beasley and Tinkham on effects of fluctuations on the
superconducting transition of tin whisker crystals \cite{[1]}.
Experimental data were compared with theories of thermally
activated phase slips, both in the linear regime and in the
non-linear regime of finite current flows.  More recently, Tinkham
and collaborators studied the loss of superconductivity in very
thin wires of MoGe, deposited on carbon nanotubes, where quantum
fluctuations are involved \cite{[2]}-\cite{[4]}. Related work from
Tinkham's laboratory, in recent years, has elucidated the
breakdown of superconductivity in ultrasmall metallic grains,
measured by the even-odd alternation of Coulomb-blockade
energies~\cite{[5],[6]} , vortex motion and resistance in high
temperature superconductors~\cite{[7],[8]} , and critical currents
in frustrated arrays of Josephson junctions.~\cite{[9]}.

The decay of supercurrents in liquid $^4$He and in Bose-Einstein
condensates of ultracold atoms has much in common with the decay
of superconductivity.  Concepts of flux-line motion, and of phase
slips due to thermal or quantum fluctuations, appear in both
cases. A new dimension has been added to the subject by recent
experimental advances, where cold atoms have been trapped in a
region that contains a spatially periodic potential, produced by
optical standing waves (see for example [\onlinecite{greiner}]).
The ability to vary continuously the parameters of the system, by
changing the strength of the periodic potential, as well as by
varying the number of trapped atoms and the shape of the overall
confining potential, allows one to explore new regimes of
parameters and to make more precise confrontations between theory
and experiment. In turn, these developments give added urgency to
the theoretical study of supercurrent decay.

In the present paper, we discuss similarities and differences
between  the decay of supercurrents in superconductors and systems
of trapped atoms, and we present some new results for the latter.
Specifically we consider certain experimental procedures which
have become standard in systems of ultra cold atoms. In the first
scheme, a condensate is prepared on a lattice with a specified
intensity, when the lattice is suddenly accelerated to a finite
velocity. In other words, a moving condensate is prepared in the
lattice frame, essentially fixing the phase gradient. A similar
experiment in superconductors would involve threading a flux
through a closed superconducting loop. Such sudden lattice boosts
were applied by several groups to demonstrate a dynamical
instability of the superfluid when the imposed phase gradient
exceeds $\pi/2$ per unit cell\cite{inguscio}. A related experiment
involves tilting the lattice, thus subjecting the atoms to a
linear potential. This is equivalent to imposing a constant
voltage on a superconductor. The technique was used, for example,
to demonstrate Bloch oscillations of a condensate\cite{salomon}.
In a third experimental sequence, one can prepare a moving
condensate, then continuously increase the depth of the lattice
toward the superfluid- Mott insulator transition.

The response of the atomic system to the perturbations,
can be measured,
by direct observation of the time evolution.
Decay of the current, for example, is observed by repeated experiments,
where atoms are released from the trap after varying waiting periods.
The phase gradient in the superfluid at the time of release may be inferred
from a {\em time of flight} measurement of the momentum distribution.
This should be contrasted with superconductors, where measurements
probe I-V characteristics.

Besides the differences in the experimental observation procedures,
there are unique features of trapped atom systems which influence the
physics of supercurrent decay.
First, to a very good approximation such systems can be
considered perfectly clean. Super-currents decay only due to breaking of
Galilean invariance by the periodic potential.

A second feature that
distinguishes the dynamics of ultra cold atoms is their nearly perfect
isolation from the environment. Strictly speaking they are always
under-damped. However, we are usually interested in the dynamics of a subset of
the system degrees of freedom, such as the super-current. How much the
dynamics of the the interesting variables is damped, depends solely on the
remaining system degrees of freedom
rather than on external dissipation sources.
In superconductors on the other hand, the order parameter is invariably
in strong contact with other degrees of freedom, and therefore its dynamics is
almost always overdamped.

\section{Critical current in the superfluid phase}

Ultra cold atoms in an optical lattice, confined to the lowest
Bloch band are described by the well known Bose Hubbard
Hamiltonan:
\be
H=-J\sum_{\langle ij\rangle}(a\yd_i a\nd_j+\mbox{H.c.})
+\frac{U}{2}\sum_i n_i(n_i-1),
\label{BHM}
\ee
where $J$ and $U$ are the hopping amplitude and the on-cite
repulsive interaction, $\langle ij\rangle$ denotes pairs of
nearest neighbor sites. Another implicit parameter in this
Hamiltonian is the average number of bosons per site, $N$. In this
paper we shall be primarily concerned with the case where $N$ is a
large integer. We shall address to separate regimes, the first
is defined by the conditions $UN^2\gg JN\gg U$, while the second
regime corresponds to the superfluid near the transition to a Mott
insulator ($UN^2\gg JN\sim U$).

If the condition $UN\gg J$ holds, then the interactions are
sufficiently strong to suppress amplitude fluctuations
of the order parameter, and
(\ref{BHM}) can be mapped to the quantum phase model:
\be
H=-JN\sum_{\av{ij}}\cos(\f_i-\f_j)-\frac{U}{2}\sum_i\left(
\frac{\partial}{\partial\f_i}\right)^2
\label{QPM}
\ee
The additional condition $JN\gg U$ ensures that the system is far
from the superfluid-insulator
transition, and facilitates a semiclassical approximation.
In the classical limit the boson creation and annihilation
operators can be treated as complex numbers subject to discrete
Gross-Pitaevskii equations~\cite{psg}:
\be
i{d\psi_j\over d t}=-J\sum_{k\in O}\psi_{k}+U|\psi_j|^2\psi_j,
\label{gp1}
\ee
where the set $O$ contains the nearest neighbors of site $j$. In
the quantum rotor limit $UN\gg J$ the number fluctuations can be
integrated out leaving us with only the equations of motion for
the phase $\phi_j=\arg \psi_j$:
\be
{d^2\phi_j\over dt^2}=-2UJN\sum_{k\in O}\sin(\phi_{k}-\phi_j).
\label{gp2}
\ee
Alternatively Eq.~(\ref{gp2}) immediately follows from the
Hamiltonian~(\ref{QPM}). Both equations (\ref{gp1}) and
(\ref{gp2}) can support stationary current carrying states
$\psi_j\propto \exp(ipx_j)$. A simple linear stability analysis
of shows~\cite{niu, bishop} that these states become unstable
towards small perturbations when the phase twist exceeds $\pi/2$
per unit cell. The onset of this instability is signaled by appearance
of imaginary frequencies.
So $\pi/2$ is the critical phase twist above which a uniform
superfluid state breaks down. This instability was recently
observed experimentally~\cite{inguscio}.

In principle, one can identify another type of instability,
characterized by appearance of negative frequencies, in systems
described by Eq.~(\ref{gp1})~\cite{niu}. In general this occurs at
a phase twist $p^\star< \pi/2$. However, in the quantum rotor
limit $UN\gg J$, where we work, the two instabilities coincide.

While the modulational instability occurs precisely at $p=\pi/2$
for $JN>>U$, we expect that the current can decay at smaller
momenta due to either quantum or thermal fluctuations (see also
Ref.~\onlinecite{pw}). We envision the following experimental
scheme to observe this. The condensate is either boosted to a
state with a certain phase gradient or gradually accelerated.
Following the boost or while the system is accelerating we probe
the evolution of the phase gradient. If the system is sufficiently
close to the modulational instability, i.e. $p$ is slightly below
$\pi/2$, the coherent motion of the condensate is expected to
decay, the larger the phase gradient, the faster this decay will
occur.

The other regime we shall address, is that of the superfluid close
to the quantum phase transition to a Mott insulator at
commensurate filling (i.e. $JN\sim U$). For simplicity we also
assume that $N\gg 1$. Since the coherence length $\xi$ diverges at
the transition, one can use a continuum semiclassical description,
after coarse-graining the system, to describe the static and
dynamic properties of the condensate. At commensurate filling the
appropriate quantum action written in terms of the superfluid
order parameter reads:~\cite{ehud, sachdev_book}
\be
S=C\int d^d x dt \left|{d\psi\over dt}\right|^2-\left|\nabla \psi\right|^2+
|\psi|^2-{1\over 2}|\psi|^4,
\label{gl}
\ee
where length is measured in units of $\xi$ and time in units of
$\xi/c$, with $c$ the sound velocity. $C$ is a numerical
prefactor. The bare parameters $\xi$, $c$, and $C$ can be found
using a mean-field approximation~\cite{pra}. For the cubic
d-dimensional lattice they read:
\be
\xi={1\over \sqrt{2d(1-u)}},\; c=2JN\sqrt{2d},\; C={1\over 2
(2d)^{d/2}}\left(1-u\right)^{3-d\over 2},
\ee
where we introduced the dimensionless interaction $u=U/U_c$ with
$U_c=8JNd$ the critical interaction strength in the mean field approximation.
The action (\ref{gl}) correctly describes low
energy dynamics of the system in the vicinity of the phase
transition, only if the couplings $\xi$, $c$, $U_c$ and $C$ are
properly renormalized. While in three dimensions the effects of
such renormalization should be weak, in two and especially one
dimensional cases they strongly modify the couplings and the
critical exponents. The bare mean field parameters can then be used
only as an estimate. Note that the
dimensionless part of the action~(\ref{gl}) is general, and so are
the conclusions we reach in this paper, once the renormalized,
rather than mean field parameters are used in the prefactors.
The action (\ref{gl})
is obviously extremized by stationary current-carrying states:
$\psi_p(x)=\sqrt{1-(p\xi)^2}\,\mathrm e^{ip x\xi}$. It is easy to
check~\cite{pra} that these states are stable with respect to
small fluctuations for  $p<p_c=1/(\xi\sqrt{3})$. Since $\xi$
diverges at the phase transition, the critical phase
twist vanishes at that point as it should.

A possible experimental procedure to measure the decay rate at low
currents follows. A condensate with a specified phase gradient
is prepared in a weak lattice (small $u$). Then, the lattice
potential is gradually increased in time, driving the system closer
to the Mott phase. This, in turn, results in the increase of the
correlation length $\xi$ and in decrease of the critical momentum
$p_c$. As $p_c$ approaches $p$ the superfluid current is expected to decay
either due to quantum or thermal fluctuations.

\section{Decay of the superfluid current}

In this section we describe how the superfluid current decays in a
lattice when $p$ is below $p_c$. We shall consider first the Gross-Pitaevskii
regime $JN>>U$ in the quantum rotor limit $UN>>J$. due to either thermal or quantum
effects.

\subsection{Gross-Pitaevskii regime}

\subsubsection{Quantum Decay}

We first discuss the decay of superfluid currents within the
quantum phase model, well away from the superfluid insulator
transition ($JN\gg U$).
The action corresponding to the quantum rotor model (\ref{QPM})
is given by
\be
S=\int d\tau\, \sum_j  {1\over 2U}\left({d\phi_j\over
d\tau}\right)^2 -\sum_{\langle j,j^\prime\rangle}
2JN\cos(\phi_{j}-\phi_{j^\prime}),
\label{s8}
\ee
or after the rescaling $\tau\to \tau/\sqrt{UNJ}$:
\be
S=\sqrt{JN\over U}s,
\label{s5}
\ee
where
\be
s= \int d\tau\, \sum_{j} {1\over 2}\left({d\phi_j\over
d\tau}\right)^2 -\sum_{\langle j,j^\prime\rangle}
2\cos(\phi_{j}-\phi_{j^\prime}).
\label{s9}
\ee
To leading order in $\sqrt{U/JN}$, which
plays the role of the effective Planck's constant for this
problem~\cite{ap1}, the tunneling rate depends on
the action $S_b$, associated with the bounce solution of the classical equations
of motion in the inverted potential~\cite{coleman}:
\be
\Gamma\propto \mathrm e^{-S_b},
\ee
Clearly the action should vanish at $p=\pi/2$, since at this
point the spectrum becomes unstable and the tunneling barrier
disappears. Deep in the superfluid regime $U/JN\ll 1$, the
tunneling is effective only if $p$ is close to $\pi/2$, where the
product $\sqrt{JN/U}s$ is not too large. In this case one can make
further progress in calculating the tunneling action by expanding
(\ref{s9}) up to cubic terms in phase differences
$\phi_{j}-\phi_{j^\prime}$:
\be
s\approx \sum_{j,{\bf k}} \int d\tau \biggl[{1\over
2}\left({d\phi_{j_{\bf k}}\over
d\tau}\right)^2+\cos(p)\,(\phi_{j+1,{\bf k}}-\phi_{j,{\bf
k}})^2+(\phi_{j,{\bf k+1}}-\phi_{j,{\bf k}})^2-{1\over
3}(\phi_{j+1,{\bf k}}-\phi_{j,\bf k})^3\biggr].
\label{s6a}
\ee
Here we explicitly split the site index into longitudinal ($j$)
and transverse (${\bf k}$) components. Also, for the convenience,
we shifted the phase $\phi_{j,{\bf k}}\to \phi_{j,{\bf k}} +p j$
so that the metsatbale state now corresponds to $\phi_{j,{\bf
k}}=0$. Note that at $p\to\pi/2$ only longitudinal modes become
soft, acquiring a prefactor $\cos p$ in front of the quadratic
term in the action. This implies that we can safely use continuum
approximation for the phases along the transverse directions. Then
instead of (\ref{s6a}) we derive:
\be
s\approx \sum_{j} \int d\tau d^{d-1} x \biggl[{1\over
2}\left({d\phi_{j}\over d\tau}\right)^2 +\left({d\phi_j\over d{\bf
x}}\right)^2+\cos(p)\,(\phi_{j+1} -\phi_{j})^2-{1\over
3}(\phi_{j+1}-\phi_{j})^3\biggr].
\label{s6b}
\ee
In this equation ${\bf x}$ denotes transverse coordinates which
reside in a $d-1$ dimensional space. Upon rescaling
\be
\phi=\cos (p) \tilde \phi,\; \tau={\tilde \tau\over
\sqrt{\cos(p)}},\; x={\tilde x \sqrt{2}\over\sqrt{\cos p}},
\label{scaling1}
\ee
the action (\ref{s6b}) simplifies further:
\be
s\approx (\pi/2-p)^{6-d\over 2}\tilde s_d,
\label{s6c}
\ee
where
\be
\tilde s_d=2^{d-1\over 2}\sum_j\int d^d\xi \, \biggl[{1\over
2}\left({d\tilde\phi_j\over d{\bf \xi}}\right)^2+
(\tilde\phi_j-\tilde\phi_{j+1})^2-{1\over
3}(\tilde\phi_j-\tilde\phi_{j+1})^3\biggr]
\ee
is just a number, which is determined only by the dimensionality
$d$. We will provide its detailed variational derivation
elsewhere~\cite{pra} and here just quote the results: $\tilde
s_1\approx 7,\; \tilde s_2\approx 25,\; \tilde s_3\approx 90$.
>From the scaling (\ref{scaling1}) it is obvious that the
characteristic transverse dimension of the instanton $x$ scales as
$(\pi/2-p)^{-1/2}\gg 1$, justifying the continuum approximation.
Above $d=6$ the tunneling action would experience a discontinuous
jump at $p=\pi/2$. However, since we deal with $d\leq 3$, the
action always continuously vanishes at $p\to\pi/2$. In this way we
derive the asymptotic decay rate of a uniform current state near
the modulation instability:
\be
\Gamma\propto \mathrm e^{-S_d}
\ee
with
\be
S_d\approx \tilde s_d\sqrt{JN\over U}(\pi/2-p)^{6-d\over 2}.
\label{sd}
\ee

\subsubsection{Thermal Decay}

To calculate the exponent, characterizing the thermal decay rate
one has to compute the difference of energies of the metastable
state and the saddle-point, which separates two adjacent
metastable minima~\cite{langer, langer1, halperin}. Both
saddle-point and metastable configurations are the stationary
solutions of the equations of motion (\ref{gp2}):
\be
\sum_{k\in O}\sin(\phi_{k}-\phi_j)=0.
\label{eq7}
\ee
The metastable state corresponds to the uniform phase twist:
$\phi_{j,{\bf k}}=j p$. The saddle-point state relevant for the
durrent decay can be easily found in one dimension:
\be
\phi_j=\left\{\begin{array}{ll} p^\prime j & j<0\\
\pi+p^\prime (j-2) & j\geq 1\end{array}\right.,
\label{eq8}
\ee
where $p^\prime\approx p-(\pi-2 p)/M$ if we use periodic boundary
conditions for the system with $M$ sites. The energy difference
between the two states in the limit $M\to\infty$ is
\be
\Delta E=2JN(2\cos p\,-\sin p\,(\pi-2p)).
\ee
Correspondingly the decay rate is proportional to
\be
\Gamma\propto \mathrm e^{-\beta\Delta E}=\mathrm
e^{-2JN\beta(2\cos p- (\pi- 2p)\sin p)}.
\label{s12}
\ee
In particular when $p\to \pi/2$ we have:
\be
\Gamma\propto\mathrm e^{-{4\over 3}NJ\beta (\pi/2-p)^3}.
\ee
In higher dimensions we cannot find the energy of the
saddle-point explicitly for all values of $p$. However, similarly
to the quantum case, at $p$ close to $\pi/2$ we can expand the
energy functional up to cubic terms in phase differences
\be
E_d\approx JN\sum_j \int d^{d-1} x \biggl[{1\over
2}\left({d\phi_j\over dx}\right)^2+\cos (p)\,
(\phi_{j+1}-\phi_j)^2-{1\over 3}(\phi_{j+1}-\phi_j)^3\biggr],
\ee
where $\phi_j(x)$ is the nontrivial solution of the corresponding
Euler-Lagrange equations vanishing at $x\to\infty$. We again
shifted the phase: $\phi_j\to \phi_j+pj$, so that $\phi_j=0$
corresponds to the current state. After rescaling
$\phi_j=\cos(p)\,\tilde\phi_j$ and $x=\tilde x \sqrt{2}/\sqrt{\cos
p}$ we find:
\bea
E_d\approx JN\, 2^{d-1\over 2}(p_c-p)^{7-d\over 2}\sum_j\int
d^{d-1}\tilde x \biggl[ {1\over 2}\left({d\tilde\phi_j\over
d\tilde x}\right)^2+(\tilde\phi_{j+1}-\tilde\phi_j)^2-{1\over
3}(\tilde\phi_{j+1}-\tilde\phi_j)^3\biggr].
\eea
Note that the integral in the expression above coincides with
$\tilde s_{d-1}$ up to a number $2^{d-2\over 2}$. So we
immediately conclude that
\bea
E_d\approx \tilde s_{d-1} JN\sqrt{2} (\pi/2-p)^{7-d\over 2}.
\eea

Note that the activation energy characterizing the thermal decay
vanishes faster as $p\to\pi/2$ than the tunneling action. It
implies that thermal fluctuations become increasingly important
and dominate the decay of superfluid current as the system
approaches the dynamical instability. Comparing the ratio $E_d/T$ and
the tunneling action (\ref{sd}) we obtain the crossover
temperature:
\be
T^\star\approx {\tilde s_{d-1}\over \tilde s_d} c \sqrt{\pi/2-p}
\label{tstar}
\ee
at which the quantum and thermal decay rates coincide. Here
$c=\sqrt{2UJN}$ is the sound speed in equilibrium (i.e. $p=0$).
Alternatively, we can fix the temperature to obtain the momentum
crossover scale $p^\star$ at which thermal and quantum decay rates
coincide:
\be
\pi/2-p^\star\approx \left({\tilde s_d\over \tilde
s_{d-1}}\right)^2 \left(T\over c\right)^2.
\ee
At phase gradients larger than $p^\star$, thermal decay dominates.
The tunneling action (\ref{sd}) at this value of
momentum is given by
\be
S_d^\star=\tilde s_d\left({\tilde s_d\over \tilde
s_{d-1}}\right)^{6-d}\sqrt{JN\over U}\left(T\over c\right)^{6-d}.
\ee
If $S_d^\star\gg 1$, then at the crossover momentum the
current decay is exponentially suppressed and will be nonzero only
at $p$ closer to $\pi/2$. Then the thermally activated phase slips
will dominate the decay process and quantum tunneling can be
ignored. In the opposite limit $S_d^\star\ll 1$ the current will
decay at $p<p^\star$ due to quantum process and the temperature
effects are unimportant. The characteristic crossover temperature
separating quantum and thermal decay regimes is thus:
\be
T_q\approx Ac\tilde s_d^{-{7-d\over 6-d}}\left({U\over
JN}\right)^{1\over 2(6-d)},
\ee
where $A$ is a numerical constant of the order of one. Note that
for all relevant dimensions $d\leq 3$ the last multiplier is
always of the order of one because of the small exponent
$1/(12-2d)$. Therefore, the crossover temperature $T_q$ is of the order of
the sound velocity (or equivalently Josephson energy).

\section{Current Decay in the vicinity of the Mott transition}

Let us now address decay of supercurrents in the regime where
$JN\sim U$ and large integer filling $N$. As we already argued, in
the vicinity of the Mott-insulator phase transition the
correlation length $\xi$ becomes large compared to the lattice
constant. One can therefore use a continuum description of the
problem~(\ref{gl}). The relativistic dynamics of (\ref{gl}) is a
special feature of the commensurate transition. Ordinary
superfluids are described by a similar action, but with a kinetic
term including first time derivatives.

The Euler Lagrange-equations derived from (\ref{gl}) admit
stationary current carrying solutions of the form
\be
\psi=\sqrt{1-(p\xi)^2}e^{ip\xi x},
\ee
We emphasize again that $x$ is measured in units of the
correlation length $\xi$. The current state becomes unstable at
$p>p_c=1/\sqrt{3\xi}$. Below $p_c$ the current can still decay due
to quantum or thermal fluctuations.

For the thermal decay only the static part of the action needs to
be considered. Then there is no difference between our problem and
the current decay in ordinary superfluids described by the
Ginzburg-Landau free energy. In particular, in context of
superconducting wires, the exponent characterizing the current
decay rate in one dimension was computed by Langer and
Ambegaokar~\cite{langer1}, and the prefactor setting the time
scale was later found by McCumber and Halperin~\cite{halperin}. In
three dimensions the asymptotic behavior of the corresponding
exponent at $p\to 0$ was obtained by Langer and Fisher~\cite{lf}.
However, here we are interested in the opposite limit $p\to p_c$.

For both the thermal and the quantum cases we will use the scaling
approach successfully applied above for the quantum phase model in
the Gross-Pitaevskii regime. We expand the action to cubic order
in the amplitude ($\h$) and phase ($\phi$) fluctuations, about the
metastable minimum, and integrate out the gapped amplitude mode.
After the rescaling:
\be
x\to\frac{x}{2\, 3^{1/4}\sqrt{\xi}\sqrt{p_c-p}},
~~z\to\frac{z}{6\xi(p_c-p)}, ~~\phi\to \phi\, {3^{3/4}\over
2}\sqrt{\xi}\sqrt{p_c-p}
\ee
the action to leading order in $p_c-p$ becomes:
\be
S=C\, {3^{9/4-d}\over 2^d}\xi^{5/2-d}(p_c-p)^{5/2-d}\int d{\bf
z}dx\, (\nabla\phi)^2+(\partial^2_x\phi)^2-(\partial_x
\phi)^3\approx A_d (1-u)^{1/4} (p_c-p)^{5/2-d}
\label{s_cont}
\ee
where $\bz$ denotes all the transverse coordinates relative to the
current direction, including time. $\nabla=({\partial_{\bf
z}},\partial_x)$ is the gradient in $d+1$ dimensions. Accordingly,
the quantum decay rate is given by $\G_{Q}\propto
\exp(-A_d(1-u)^{1/4}(p_c-p)^{5/2-d})$. A variational
calculation\cite{pra}, yields $A_1\approx 18.4$ and $A_2\approx
8.4$. As before, to calculate the thermal decay rate one simply
has to substitute $d\to d-1$, so that
\be
\G_{T}(d)\propto \exp\left(-\frac{JN}{T}(2d)^{-3/4}(1-u)^{1/4}
A_{d-1}(p_c-p)^{7/2-d}\right).
\ee
In the one-dimensional case the relevant constant $A_0\approx
12.56$. It is interesting to contrast these results with the
asymptotic decay rate (\ref{sd}), found in the Gross-Pitaevskii
regime. First we observe that the tunnelling action close to the
Mott insulator vanishes as a smaller power of $p_c-p$. Moreover,
for $d=3$ the scaling hypothesis for the quantum decay rate breaks
down, suggesting that $S$ is discontinuous at the critical current
and is dominated by the fluctuations of a finite (not diverging as
$p\to p_c$) length scale. We therefore expect that in three
dimensions at zero temperature the instability marks a sharp
localization transition. At finite $T$, thermal fluctuations
broaden this transition, because the activation barrier energy
vanishes at $p_c$ for $d<7/2$.

The quantum to thermal crossover for a given dimensionless
interaction and phase gradient is found by comparing the two decay
rates. In one and two dimensions we find
\be
T^\star(p)=\frac{JN}{(2d)^{3/4}}\frac{A_{d-1}}{A_d}(p_c-p).
\label{Tstarp}
\ee
In three dimensions $T^\star=0$ because the quantum decay is
effectively suppressed. As discussed above for the quantum phase
model, there is a more useful $p$ independent, crossover
temperature scale. Using the same arguments as in the
Gross-Pitaevskii we can find the temperature separating the
quantum and thermal decay regimes:
\be
T_q\sim JN A_{d-1} A_d^{-{7-2d\over 5-2d}}(1-u)^{-{1\over 10-4d}}.
\ee
We see that near the Mott transition the crossover temperature
strongly depends on interaction $u$. Thus as $u\to 1$ then in one
and two dimensions $T_q\to\infty$ so the quantum decay always
dominates over the thermal. In particular in two dimensions we
find $T_q\sim 0.03 JN/\sqrt{1-u}$ and in the one dimensional case
$T_q\sim 0.1 JN /(1-u)^{1/6}$, i.e. the crossover temperature is
very high and the thermal decay is unimportant.

\section{Dynamics of the Current Decay}

Except for the zero temperature decay in the continuum
relativistic model in three dimensions, there is no sharp
transition between superfluid current carrying state and the
insulating state with no current. Indeed, in all other cases the
tunneling action and the energy barrier continously vanish as the
system approaches the point of the modulation instability. Thus,
instead of a sharp transition boundary we can define a broad
crossover region, say $1<S_d<3$, which separates the superfluid
phase with a relatively slow current decay and the insulating
phase with a fast decay. The fact that the transition is broad
does not imply, however, that within a single experiment a gradual
current decay will be detected as the system slowly tuned through
the crossover region. The tunneling and thermal decay rate define
a probability of creating a single phase slip per lattice site. In
the following evolution the phase slip can either rapidly release
its energy into phonon (Bogoliubov's) modes and bring the system
to a next metastable minimum with a lower current, or it can
trigger the current decay in the whole system.

\begin{figure}[ht]
\includegraphics[width=15cm]{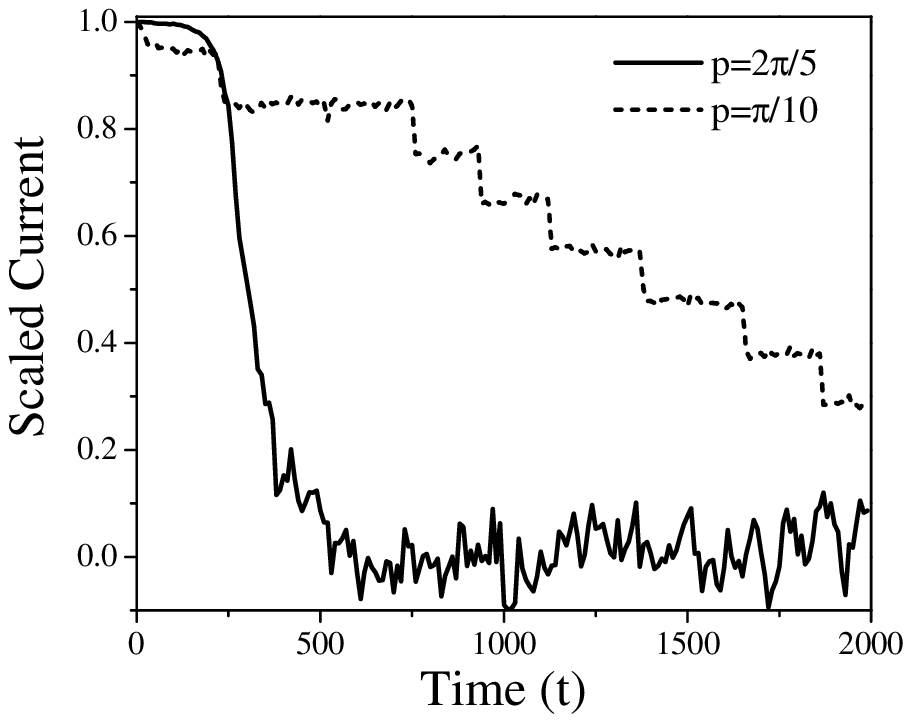}
\caption{Current (scaled to one at $t=0$) versus time for a one
dimensional periodic array of $200$ sites with two different
initial phase gradianets. The evolution is determined solving
equations of motion (\ref{gp1}) with constant hopping amplitude
$J=1$ and interaction increasing in time $U=0.01 \tanh 0.01 t$ for
$p=2\pi/5$ and $U=\tanh 0.01 t$ for $p=\pi/10$. To get the current
decay we add small fluctuations to the initial values of the
classical fields $\psi_j(t=0)$.}
\label{fig_damp}
\end{figure}

In a closed system, i.e. with no coupling to the environment,
these two regimes are well defined because the damping of the
phase slip comes from the internal degrees of freedom, which are
completely described by the equations of motion. Also near the
critical current there should be very little difference between
the overdamped and underdamped regimes in the thermal and quantum
cases. This is because the energy barrier is very small and the
classically allowed motion after the tunneling event starts very
close to the metsatable maximum. To see which of the regimes is
realized in our systems we numerically solve the classical
(Gross-Pitaveskii) equations of motion (\ref{gp1}). We start from
a uniform current state in a periodic lattice. To allow for a
current decay we add small fluctuations to the initial values of
the classical fields $\psi_j(t=0)$. This mimics the effect of
thermal fluictuations. In Fig.~\ref{fig_damp} we plot the computed
current versus time for a one-dimensional array of $M=200$ sites.
Initially the system is asumed to be noninteracting ($U=0$)
macroscopically occupied state with a given phase gradient $p$
(specifically we consider $p=2\pi/5$ and $p=\pi/10$) and unit
hopping. Then interaction is gradually increasing in time and
reaches some constant value and the we follow the time evolution
of the current. It is clear from the figure that the phase slips
in the smaller current case ($p=\pi/10$) are overdamped leading to
gradual decay. With the used parameters the total winding number
is $200\pi/10/2\pi=10$ and the figure clearly shows that each
phase slip decreases the current by roughly ten percent. On the
other hand for the larger current ($p=2\pi/5$) a single phase slip
generates immediate current decay in the whole sample and this
corresponds to the underdamped regime. We will not attempt here to
find the precise boundary between the scenarios, since it is not
the purpose of this paper. We would like to stress that near the
instability the system is always in the underdamped regime. We
checked that a similar overdamped to underdamped crossover occurs
in other spatial dimensions. So if $p$ is not too small, the in a
given experiment one will always see a sharp transition from the
superfluid to the insulating regime. However the precise point,
where the current decays will depend on the details of the
experiment, for example on the rate of change of external
parameters like tunneling, interaction, or the phase gradient $p$
if the system is accelerated. On the other hand in the absence of
any fluctuations the transition is very sharp and always occurs at
$p=\pi/2$.

We can do a similar numerical analysis at small currents close to
the Mott transition using the action~(\ref{gl}). We find that even
close to the critical current the decay is overdamped, i.e. the
current decays in steps. This result is natural because the size
of the phase slip in this case is large and hence it is very easy
to dissipate the released energy into phonon modes.

We would like to mention that if the motion of phase slips is
underdamped then in a truly infinite system the current state is
always unstable. Indeed the probability of a phase slip in the
whole system is proportional to its size $M$. If this single phase
slip causes triggers the current decay in the whole sample, then
obviously a state with a uniform phase gradient cannot exist.
However, in finite size systems these effects are not so crucial,
because the decay probability depends exponentially on the
couplings and current but only linearly in a system size.

\section{Conclusions}

The modulational instability can be observed either by increasing
momentum of the superfluid accelerating the condensate or by
increasing lattice potential and driving the system closer to the
Mott transition, while the condensate is in motion. We showed that
because of quantum or thermal effects the current decays before
the system becomes classically unstable. Therefore instead of a
sharp transition, there is a crossover region where the decay rate
grows from being exponentially small to large. The crossover
region becomes narrower either deep in the superfluid regime (i.e.
$JN\gg U$) or in higher dimensions. In particular, in the
three-dimensional case we always expect a very sharp boundary
separating the regions with very weak and very strong decay rates.

We found that deep in the superfluid regime the crossover
temperature separating quantum and thermal decay is of the order
of the Josephson plasma frequency in all dimensions. At small
currents, close to the Mott phase, the decay occurs through
thermal fluctuations in three dimensions and through quantum
tunneling in one and two dimensions.

We argue that both overdamped and underdamped dynamics of the
current decay can be realized in these systems. The underdamped
regime corresponds to high currents close to $p=\pi/2$, while at
low currents the dynamics is overdamped.

\acknowledgements

We would like to acknowledge useful discussions with D.-W. Wang.
This work was supported by ARO and NSF under grants DMR-0233773,
DMR-0231631, DMR-0213805.

\newpage

\end{document}